\documentclass[10pt,conference]{IEEEtran}

\usepackage{cite}
\usepackage{amsmath,amssymb,amsfonts}
\usepackage{algorithmic}
\usepackage{graphicx}
\usepackage{textcomp}
\usepackage[usenames,dvipsnames]{xcolor}
\usepackage{url}
\usepackage[hidelinks,bookmarks=false]{hyperref}

\usepackage{listings}
\usepackage{fancyvrb}
\usepackage{booktabs}
\usepackage[absolute]{textpos}

\makeatletter
\def\endthebibliography{%
	\def\@noitemerr{\@latex@warning{Empty `thebibliography' environment}}%
	\endlist
}

\makeatother

\begin{document}

\title{Towards a Knowledge Base of Common Sustainability Weaknesses in Green Software Development\vspace{-0.4em}}

\author{
	\IEEEauthorblockN{Priyavanshi Pathania$^\dagger$, Rohit Mehra$^\dagger$, Vibhu Saujanya Sharma$^\dagger$, Vikrant Kaulgud$^\dagger$, Sanjay Podder$^\ddagger$, Adam P. Burden*}
	\IEEEauthorblockA{$^\dagger$Accenture Labs, India | $^\ddagger$Technology Sustainability Innovation, Accenture, India | *Accenture, USA\\
		\{priyavanshi.pathania, rohit.a.mehra, vibhu.sharma, vikrant.kaulgud, sanjay.podder, adam.p.burden\}@accenture.com\vspace{-0.2em}}
}

\maketitle

\begin{textblock*}{10cm}(1.5cm,26.2cm) 
	DOI: \url{https://doi.org/10.1109/ASE56229.2023.00204}
\end{textblock*}

\begin{abstract}

With the climate crisis looming, engineering sustainable software systems become crucial to optimize resource utilization, minimize environmental impact, and foster a greener, more resilient digital ecosystem. For developers, getting access to automated tools that analyze code and suggest sustainability-related optimizations becomes extremely important from a learning and implementation perspective. However, there is currently a dearth of such tools due to the lack of standardized knowledge, which serves as the foundation of these tools. In this paper, we motivate the need for the development of a standard knowledge base of commonly occurring sustainability weaknesses in code, and propose an initial way of doing that. Furthermore, through preliminary experiments, we demonstrate why existing knowledge regarding software weaknesses cannot be re-tagged \textit{``as is''} to sustainability without significant due diligence, thereby urging further explorations in this ecologically significant domain.

\end{abstract}


\section{Introduction}\label{introduction}

While software serves as the backbone of virtually all intelligent solutions designed to support environmental sustainability, it has a carbon footprint of its own. The internet and communications technology industry, which encompasses software and the corresponding hardware infrastructure, currently accounts for 2-7\% of global greenhouse gas emissions \cite{unepccc, FREITAG2021100340}. With compute-intensive technologies like metaverse, generative artificial intelligence, blockchain, etc. gaining more importance, these emissions are predicted to increase to a massive 14\% by 2040 \cite{unepccc}. Failure, or even a delay in addressing these rapidly growing emissions can have a catastrophic impact on the long-term sustainability of our environment.

One of the primary reasons behind these high carbon emissions is the non-optimization of software from a sustainability perspective \cite{10.1145/3154384}. There are multiple inherent reasons like lack of awareness, lack of encouragement, hidden impact, etc. attributing to this non-optimization \cite{9793921, 10.1145/3350768.3350770, 7169668}. However, we believe that the dearth of standard sustainability-related knowledge, especially around commonly occurring sustainability weaknesses and their corresponding energy impact, is still one of the foremost \cite{7169668, 8449016, 10.1145/1387830.1387835}. We define a sustainability weakness as a part of code that has a detrimental impact on energy consumption and can potentially be remediated by leveraging an energy-efficient alternative. This limitation further cascades into (i) lack of automated tools that can detect such sustainability weaknesses in code, gauge their impact, and offer potential remediation strategies (ii) lack of learning material to educate the relevant stakeholders (developers, testers, etc.) on how to avoid/mitigate these weaknesses (iii) lack of training datasets to train the next generation of code generation artificial intelligence models for generating sustainable code (iv) and lack of approaches to certify a software as environment-friendly.

Currently, multiple standards/knowledge bases exist within the software engineering community that provides visibility into common weaknesses across multiple other dimensions. For example, the Common Weakness Enumeration (CWE) is a popular community-developed standard of commonly occurring software and hardware weaknesses \cite{cwe}. More focused standards include the ISO-5055 \cite{9734273} and ASCQM \cite{ascqm} concerning software quality attributes like performance efficiency, reliability, and maintainability, OWASP \cite{wichers2013owasp} for web application security, among others. Most of these standards are either built on top of other existing standards (for example, ISO-5055 and ASCQM using CWE) or have eventually been cross-referenced with each other. Moreover, weaknesses as part of these standards are systematically collated by the open community/subject-matter-experts, are well studied for their impact, and can act as a baseline for weakness identification, mitigation, and prevention efforts. Unfortunately, to the best of our knowledge, no such standard currently exists for software sustainability weaknesses, and development of such a standard is the key position of this paper.

Similar to other focused standards, we believe that a good starting point for systematically collating sustainability weaknesses is to study the already cataloged weaknesses in other dimensions like security, performance, and reliability for their impact on sustainability. If proven detrimental, a weakness can then be categorized as a sustainability weakness. Since CWE is a well-referenced knowledge base that encompasses weaknesses across multiple dimensions and currently includes over 932 enumerated weaknesses, it can serve as an ideal initial candidate for such a study. While a few recent works have attempted to explore this area in general, their investigations have been ad-hoc, focusing only on limited weaknesses across different granularities, such as in the context of code smells \cite{7884614, 9240625}, and remain largely insufficient to be leveraged for developing a systematic and comprehensive knowledge base.


\begin{table*}[t]
	\ttfamily
	\huge
	\centering
	\caption{Results from our early experiments to study the categorization of existing weaknesses as a sustainability weaknesses.}
	\label{tab:table_results}
	\vspace{-0.5em}
	\resizebox{\textwidth}{!}{%
	\begin{tabular}{@{}llccccll@{}}
		\toprule
		&                                        & \multicolumn{2}{c}{\textbf{Unoptimized Code}}              & \multicolumn{2}{c}{\textbf{Optimized Code}}                 &                                                                                 &                                                         \\ \cmidrule(lr){3-4} \cmidrule(lr){5-6}
		\textbf{}           & \textbf{Reference} & \textbf{Execution Time (s)} & \textbf{Energy Consumed (J)} & \textbf{Execution Time (s)} & \textbf{Energy Consumed (J)}  & \textbf{Energy Impact (\% Diff.)}                                               & \textbf{Sustainability Weakness?}       \\ \midrule
		\textbf{Weakness 1} & CWE-1046           & 41.43                       & 285.49                       & \hspace{4.0 mm}10.18        & \hspace{4.0 mm}70.97          & \multicolumn{1}{c}{\textbf{\textcolor{ForestGreen}{- \hspace{4.0 mm}75.14 \%}}} & \multicolumn{1}{c}{\textbf{True}}       \\
		\textbf{Weakness 2} & CWE-595            & 27.73                       & 192.99                       & 114.30                      & 795.56                        & \multicolumn{1}{c}{\textbf{\textcolor{red}{+ 312.23 \%}}}                       & \multicolumn{1}{c}{\textbf{\hspace{4.0 mm}False}}                       \\ \bottomrule		
	\end{tabular}
}
\vspace{-0.3 em}
\end{table*}

In the next section, we demonstrate the impact of selected CWE weaknesses on software sustainability and assess the feasibility of classifying them as sustainability weaknesses.
\section{Categorizing Existing Weakness As Sustainability Weakness}\label{experiments}

We hypothesize that not all existing weaknesses will impact sustainability. However, among those that do, they can have either a positive or a negative impact, and only the weaknesses with a negative impact can be categorized as sustainability weaknesses. Therefore, to confirm our hypothesis and study the sustainability impact of existing weaknesses, we conducted a set of early experiments.

For our experiments, we selected two existing weaknesses from CWE that have previously been studied to be among the top weaknesses that are most frequently fixed by developers \cite{MARCILIO2020110671}, and have also been included in other focused knowledge bases. The code snippets for the experiments were developed in Java, using the example code snippets provided by CWE and SonarQube (a widely used static code analysis tool) \cite{sonarqube}. Please note, that the weakness can appear to be reworded depending on the source and specific instantiation, and hence we performed a manual semantic match to correlate the same weakness across sources. For both the weaknesses, a pair of unoptimized (weakness included) and optimized (weakness remediated) code snippets were developed. Moreover, to test the correctness of these code snippets, they were statically analyzed via SonarQube to confirm the existence of weaknesses. The experiments were conducted on an m5.xlarge \cite{aws} virtual machine instance (Amazon AWS) with an Intel Xeon Platinum 8259 CL as the underlying processor, and the energy consumption was estimated using a combination of psutil \cite{psutil} and ESAVE \cite{esave}. Finally, due to the code snippets being very small, they were executed a million times to exercise significant execution time and energy consumption. Following are the two experiments that were conducted:

\subsubsection{\textbf{CWE 1046 - Creation of immutable text using string concatenation}}

Apart from being categorized under CWE \cite{cwe1046}, this weakness is also categorized as a performance efficiency weakness in ASCQM and is further described/instantiated as ``\textit{Strings should not be concatenated using + in a loop}" by Marcilio et. al. \cite{MARCILIO2020110671} and SonarQube \cite{sonarqube1046}. The presence of this weakness indicates a performance bottleneck within a loop. Java strings are immutable, hence every time they are manipulated in a loop, a new object is created in the heap memory, which consumes more memory and slows down the program execution. The recommended mitigation strategy is to leverage a mutable text buffer data element like the StringBuilder \cite{stringbuilder} or StringBuffer \cite{stringbuffer} for better performance efficiency. Table \ref*{tab:table_results} highlights that the optimized code leads to a 75.14\% reduction in energy consumption, as compared to the unoptimized variant. As per the results, it will be safe to categorize this weakness, as a sustainability weakness.

\subsubsection{\textbf{CWE 595 - Comparison of object references instead of object contents}}

Apart from being categorized under CWE \cite{cwe595}, this weakness is also categorized as a reliability weakness in ASCQM and is further described/instantiated as ``\textit{Strings and boxed types should be compared using equals()}" by Marcilio et. al. and SonarQube \cite{sonarqube595}. The rationale behind the weakness is that `==' or `!=' operators do not perform semantic equality but only check for same object references, and should only be used to compare primitive types or object references. The recommended mitigation strategy is to use the equals() \cite{equals} method for comparing two objects. Table \ref*{tab:table_results} highlights that the optimized code leads to a 312.23\% increase in energy consumption, as compared to the unoptimized variant. As per the results, it will not be safe to categorize this weakness, as a sustainability weakness. While we acknowledge, that the weakness might still need to be addressed from a reliability standpoint, it should not be a concern while evaluating/certifying a software system for sustainability.

Despite being early, our initial experiments highlight the need for a systematic and in-depth analysis of existing weaknesses, to evaluate their impact on sustainability, and thereafter, if feasible categorize them as sustainability weaknesses. A systematic and comprehensive analysis will enable the development of a knowledge base that various tools and approaches can thereafter be built on top of. Moreover, analyzing existing weaknesses will only be one of the steps toward achieving this vision. Other important steps will include (i) creating a standard evaluation framework for analyzing these weaknesses in a replicable and reproducible manner (ii) creating standard rules for tagging a weakness as a sustainability weakness (iii) creating a repository of sustainability weaknesses and their corresponding energy impacts across different experimentations (iv) and creating relevant prioritization and ranking mechanisms for arranging the weaknesses in a logical order, among others. Further explorations in these directions represent a major portion of our planned future work.
\section{Conclusions}\label{conclusion}

In this paper, we emphasized the importance of a standardized knowledge base for sustainability weaknesses and proposed leveraging CWE and similar standards for this purpose. We also highlighted the challenges associated with re-tagging existing software weaknesses to sustainability. Acknowledging the need for further exploration, we aim to encourage the software engineering community to join forces in a focused and collaborative effort to achieve our shared green vision.

\section*{Acknowledgement}

The authors would like to thank Sanjay Mittal, Ravi Kiran Velama, and Ramya Narasimhan for their generous support.

\bibliographystyle{IEEEtran}
\bibliography{Bibliography} 

\begin{thebibliography}{10}
\providecommand{\url}[1]{#1}
\csname url@samestyle\endcsname
\providecommand{\newblock}{\relax}
\providecommand{\bibinfo}[2]{#2}
\providecommand{\BIBentrySTDinterwordspacing}{\spaceskip=0pt\relax}
\providecommand{\BIBentryALTinterwordstretchfactor}{4}
\providecommand{\BIBentryALTinterwordspacing}{\spaceskip=\fontdimen2\font plus
\BIBentryALTinterwordstretchfactor\fontdimen3\font minus
  \fontdimen4\font\relax}
\providecommand{\BIBforeignlanguage}[2]{{%
\expandafter\ifx\csname l@#1\endcsname\relax
\typeout{** WARNING: IEEEtran.bst: No hyphenation pattern has been}%
\typeout{** loaded for the language `#1'. Using the pattern for}%
\typeout{** the default language instead.}%
\else
\language=\csname l@#1\endcsname
\fi
#2}}
\providecommand{\BIBdecl}{\relax}
\BIBdecl

\bibitem{unepccc}
United-Nations-Environment-Programme-Copenhagen-Climate-Centre, ``Greenhouse
  gas emissions in the ict sector: trends and methodologies,''
  \url{https://c2e2.unepccc.org/wp-content/uploads/sites/3/2020/03/greenhouse-gas-emissions-in-the-ict-sector.pdf},
  Accessed - 10/05/2023.

\bibitem{FREITAG2021100340}
\BIBentryALTinterwordspacing
C.~Freitag, M.~Berners-Lee, K.~Widdicks, B.~Knowles, G.~S. Blair, and
  A.~Friday, ``The real climate and transformative impact of ict: A critique of
  estimates, trends, and regulations,'' \emph{Patterns}, vol.~2, no.~9, p.
  100340, 2021. [Online]. Available:
  \url{https://www.sciencedirect.com/science/article/pii/S2666389921001884}
\BIBentrySTDinterwordspacing

\bibitem{10.1145/3154384}
\BIBentryALTinterwordspacing
G.~Pinto and F.~Castor, ``Energy efficiency: A new concern for application
  software developers,'' \emph{Commun. ACM}, vol.~60, no.~12, p. 68–75, nov
  2017. [Online]. Available: \url{https://doi.org/10.1145/3154384}
\BIBentrySTDinterwordspacing

\bibitem{9793921}
R.~Mehra, V.~S. Sharma, V.~Kaulgud, S.~Podder, and A.~P. Burden, ``Towards a
  green quotient for software projects,'' in \emph{2022 IEEE/ACM 44th
  International Conference on Software Engineering: Software Engineering in
  Practice (ICSE-SEIP)}, 2022, pp. 295--296.

\bibitem{10.1145/3350768.3350770}
\BIBentryALTinterwordspacing
L.~Karita, B.~C. Mour\~{a}o, and I.~Machado, ``Software industry awareness on
  green and sustainable software engineering: A state-of-the-practice survey,''
  in \emph{Proceedings of the XXXIII Brazilian Symposium on Software
  Engineering}, ser. SBES '19.\hskip 1em plus 0.5em minus 0.4em\relax New York,
  NY, USA: Association for Computing Machinery, 2019, p. 501–510. [Online].
  Available: \url{https://doi.org/10.1145/3350768.3350770}
\BIBentrySTDinterwordspacing

\bibitem{7169668}
P.~Lago, ``Challenges and opportunities for sustainable software,'' in
  \emph{2015 IEEE/ACM 5th International Workshop on Product Line Approaches in
  Software Engineering}, 2015, pp. 1--2.

\bibitem{8449016}
C.~Haupt, M.~Meinel, and T.~Schlauch, ``The software engineering initiative of
  dlr: Overcome the obstacles and develop sustainable software,'' in \emph{2018
  IEEE/ACM 13th International Workshop on Software Engineering for Science
  (SE4Science)}, 2018, pp. 16--19.

\bibitem{10.1145/1387830.1387835}
\BIBentryALTinterwordspacing
R.~A. Martin and S.~Barnum, ``Common weakness enumeration (cwe) status
  update,'' \emph{Ada Lett.}, vol. XXVIII, no.~1, p. 88–91, apr 2008.
  [Online]. Available: \url{https://doi.org/10.1145/1387830.1387835}
\BIBentrySTDinterwordspacing

\bibitem{cwe}
\BIBentryALTinterwordspacing
R.~A. Martin and S.~Barnum, ``Common weakness enumeration (cwe) status
  update,'' \emph{Ada Lett.}, vol. XXVIII, no.~1, p. 88–91, apr 2008.
  [Online]. Available: \url{https://doi.org/10.1145/1387830.1387835}
\BIBentrySTDinterwordspacing

\bibitem{9734273}
B.~Curtis, R.~A. Martin, and P.-E. Douziech, ``Measuring the structural quality
  of software systems,'' \emph{Computer}, vol.~55, no.~3, pp. 87--90, 2022.

\bibitem{ascqm}
O.~S.~D. Organization, ``Ascqm - automated source code quality measures,''
  \url{https://www.omg.org/spec/ASCQM/}, Accessed - 10/05/2023.

\bibitem{wichers2013owasp}
D.~Wichers, ``Owasp top-10 2013,'' \emph{OWASP Foundation, February}, vol.~12,
  2013.

\bibitem{7884614}
A.~Carette, M.~A.~A. Younes, G.~Hecht, N.~Moha, and R.~Rouvoy, ``Investigating
  the energy impact of android smells,'' in \emph{2017 IEEE 24th International
  Conference on Software Analysis, Evolution and Reengineering (SANER)}, 2017,
  pp. 115--126.

\bibitem{9240625}
G.~Catolino, ``A blessing in disguise? assessing the relationship between code
  smells and sustainability,'' in \emph{2020 IEEE International Conference on
  Software Maintenance and Evolution (ICSME)}, 2020, pp. 779--780.

\bibitem{MARCILIO2020110671}
\BIBentryALTinterwordspacing
D.~Marcilio, C.~A. Furia, R.~Bonifácio, and G.~Pinto, ``Spongebugs:
  Automatically generating fix suggestions in response to static code analysis
  warnings,'' \emph{Journal of Systems and Software}, vol. 168, p. 110671,
  2020. [Online]. Available:
  \url{https://www.sciencedirect.com/science/article/pii/S016412122030128X}
\BIBentrySTDinterwordspacing

\bibitem{sonarqube}
sonarsource.com, ``Sonarqube,''
  \url{https://www.sonarsource.com/products/sonarqube/}, Accessed - 17/05/2023.

\bibitem{aws}
Amazon, ``Amazon ec2 m5 instances,''
  \url{https://aws.amazon.com/ec2/instance-types/m5/}, Accessed - 10/05/2023.

\bibitem{psutil}
pypi.org, ``psutil,'' \url{https://pypi.org/project/psutil/}, Accessed -
  10/05/2023.

\bibitem{esave}
\BIBentryALTinterwordspacing
P.~Pathania, R.~Mehra, V.~S. Sharma, V.~Kaulgud, S.~Podder, and A.~P. Burden,
  ``Esave: Estimating server and virtual machine energy,'' in \emph{Proceedings
  of the 37th IEEE/ACM International Conference on Automated Software
  Engineering}, ser. ASE '22.\hskip 1em plus 0.5em minus 0.4em\relax New York,
  NY, USA: Association for Computing Machinery, 2023. [Online]. Available:
  \url{https://doi.org/10.1145/3551349.3561170}
\BIBentrySTDinterwordspacing

\bibitem{cwe1046}
CWE, ``Cwe-1046: Creation of immutable text using string concatenation,''
  \url{https://cwe.mitre.org/data/definitions/1046.html}, Accessed -
  10/05/2023.

\bibitem{sonarqube1046}
sonarsource.com, ``Strings should not be concatenated using '+' in a loop,''
  \url{https://rules.sonarsource.com/java/type/Code%20Smell/RSPEC-1643},
  Accessed - 10/05/2023.

\bibitem{stringbuilder}
Oracle, ``Class stringbuilder,''
  \url{https://docs.oracle.com/javase/8/docs/api/java/lang/StringBuilder.html},
  Accessed - 10/05/2023.

\bibitem{stringbuffer}
Oracle, ``Class stringbuffer,''
  \url{https://docs.oracle.com/javase/8/docs/api/java/lang/StringBuffer.html},
  Accessed - 10/05/2023.

\bibitem{cwe595}
CWE, ``Cwe-595: Comparison of object references instead of object contents,''
  \url{https://cwe.mitre.org/data/definitions/595.html}, Accessed - 10/05/2023.

\bibitem{sonarqube595}
sonarsource.com, ``"==" and "!=" should not be used when "equals" is
  overridden,'' \url{https://rules.sonarsource.com/java/type/Bug/RSPEC-4973},
  Accessed - 10/05/2023.

\bibitem{equals}
Oracle, ``Class string,''
  \url{https://docs.oracle.com/javase/8/docs/api/java/lang/String.html#equals-java.lang.Object-},
  Accessed - 10/05/2023.

\end{thebibliography}

\end{document}